\def\PRL#1#2#3{{\sl Phys. Rev. Lett.} {\bf #1} ~(#3) #2 }

\def\PLB#1#2#3{{\sl Phys. Lett.} {\bf B#1}~(#3) #2}
\def\PREP#1#2#3{{\sl Phys. Rep.} {\bf #1}~ (#3) #2 }
\def\NPB#1#2#3{{\sl Nucl. Phys.} {\bf B#1} ~(#3) #2 }

\def\beq{\begin{equation}}
\def\eeq{\end{equation}}
\def\bea{\begin{eqnarray}}
\def\eea{\end{eqnarray}}
\def\bq{\begin{quote}}
\def\eq{\end{quote}}

\def \gsim{\mathrel{\vcenter
     {\hbox{$>$}\nointerlineskip\hbox{$\sim$}}}}

\def\gappeq{\mathrel{\rlap {\raise.5ex\hbox{$>$}}
{\lower.5ex\hbox{$\sim$}}}}

\def\lappeq{\mathrel{\rlap{\raise.5ex\hbox{$<$}}
{\lower.5ex\hbox{$\sim$}}}}

\def\bbz{fa Z \kern-8.9pt Z}
\documentstyle [12pt,axodraw]{article}
\begin{document}
\thispagestyle{empty}
\vspace*{-1cm}
\begin{flushright}
{CERN-TH/97-14} \\
{MPI-PhT-97-004} \\
{hep-ph/9702247} \\
\end{flushright}
\vspace{1cm}
\begin{center}
{\large {\bf Flavour-Dependent and Basis-Independent 
Measures of $R$ Violation}} \\
\vspace{.2cm}
\end{center}

\begin{center}
{\bf Sacha Davidson} \\
\vspace{.3cm}
{Max Planck Institut fur Physik,
 F\"ohringer Ring 6, 80805 M\"unchen, Germany} \\

\vspace{.3cm}
and\\
\vspace{.3cm}
{\bf John Ellis} \\
\vspace{.3cm}
{ Theory Division, CERN, CH-1211, Gen\`eve 23, Switzerland} \\
\end{center}

\vspace{2cm}

\begin{abstract}

We construct lepton flavour-dependent and basis-independent 
measures of $R$-parity violation in the minimal
supersymmetric extension of the Standard Model (MSSM).
We work in the context of exact supersymmetry, neglecting
the effects of Higgs vacuum expectation values and 
soft supersymmetry-breaking terms. We devote particular attention to
appropriate choices of flavour eigenstates, and to the counting
and enumeration of $R$-violating invariants in two- and three-generation
models. We also make an indicative application of our results
to derive possible basis-independent cosmological upper bounds
on flavour-dependent violation of $R$ parity.

\end{abstract}
\vspace{3.5cm}
\begin{flushleft}
CERN-TH/97-14\\
February 1997\\
\end{flushleft}

\newpage

\section{Introduction}

One of the most attractive possible extensions of the
Standard Model is supersymmetry (SUSY) \cite{SUSY}.
The Lagrangian for the minimal supersymmetric extension
of the Standard Model (MSSM) could in principle
contain renormalisable baryon- ($B-$)  and lepton- ($L-$) 
number-violating interactions. These are constrained by
the fact that neither $B$ nor $L$ violation has yet
been observed experimentally. It is possible to remove
these interactions by imposing a 
multiplicatively-conserved global symmetry called
$R$ parity \cite{RPth}, whose action can be defined 
at the level of the MSSM superpotential such that matter superfields
($L,Q,E^c,U^c, D^c$) 
are multiplied by $-1$, but the Higgs superfields are invariant. In
component notation, $R$ parity multiplies
fields by $(-1)^{3B + L + 2S}$, where $S$ is the spin.

Alternatively, one can merely require that possible $R$-violating
couplings be sufficiently small that their
effects would not (yet) have been seen by experiment~\cite{RPexpt}.
One might also want to bear in mind the drastic cosmological
implications of possible $R$ violation, even at levels below
the current experimental upper limits, for example for the
survival of any primordial baryon or lepton asymmetry \cite{CDEO,DR}.
However, there is problem with the formulation of
experimental or cosmological
upper bounds on the  magnitudes of $R$-violating couplings:
$L$ violation can be moved around the superpotential by basis 
transformations, so the apparent magnitudes of these coupling constants
are basis dependent. In a previous paper~\cite{DE} we have 
therefore derived basis-independent expressions that
parametrize $R$ violation in the MSSM: we denote~\cite{DE} subsequently
as DE. 

The $R$-violating invariants of DE have various limitations: for
example, there we neglected soft
SUSY-breaking interactions and
the Higgs vacuum expectation values, which we do here also. 
Another shortcoming of DE was that
we summed over lepton
flavours, since otherwise the invariants would have
depended on the basis chosen for the singlet right-handed leptons.
This means that the basis-independent expressions of DE only measured
the sum of the violation of the lepton flavours
$ L_e$, $L_{\mu}$ and $ L_{\tau}$.
In this paper, we determine the appropriate
basis for the quarks and singlet right-handed leptons,
obviating the need to sum over
the quark and singlet lepton flavours.
In this way, we obtain
basis-independent measures of the violation of individual lepton flavours.
Since there are more terms in the flavour sums of DE than there
are $R$-violating couplings,
not all the terms in our previous  sums 
can be independent invariants.
We therefore discuss the number of distinct,
basis-independent combinations of coupling constants
that measure $R$-parity breaking
if one does {\it not} sum over  lepton flavours,
and express the ``superfluous'' invariants in terms of an independent
basis set. We then use these to extend our previous discussion of
the possible cosmological constraints on
$R$-parity violation to include the case where
additional flavour symmetries are present.

\section{Review of Flavour- and Basis-Independent Measures of $R$
Violation}

The problem that DE sought to address is that
the magnitude of possible $R$-violating superpotential
couplings is basis dependent.
The MSSM superpotential, including possible $R$-violating terms, is:
\beq
W = \mu_H \bar{H} H + h_u^{ij} \bar{H} Q_i U^c_j + 
 h_d^{ij} H Q_i D^c_j + h_e^{ij} H L_i E^c_j 
 +\epsilon_i \bar{H}  L_i + \lambda_1^{ijk} L_iL_j E^c_k + 
 \lambda_2^{ijk} L_iQ_jD^c_k + \lambda_3^{ijk} U^c_iD^c_jD^c_k  
\label{S1}
\eeq
Hereon, we focus on the violation of lepton number, and so neglect
the possible $B$-violating coupling constant $\lambda_3$.
It is well known that  $\epsilon$ in (\ref{S1}) can be rotated away by a 
basis transformation. This is because 
the doublet Higgs and lepton
superfields $H, L_i$ have the same gauge quantum numbers,
and SUSY removes the Standard Model 
distinction between $H$ and the $L_i$ that is  based on their spins.
Once we allow for $R$
violation, there is no unambiguous distinction between
$H$ and the $L_i$. If it is unclear how to define
a lepton, then
it is also unclear which coupling constants violate $L$.

If the Higgs and doublet lepton superfields are assembled
into a vector
\beq
\phi_I = (H, L_i) \label{phi}
\eeq
whose index $I$ runs from 0 to 3, then the superpotential
(\ref{S1}) can be rewritten as:
\beq
W= \mu^I \bar{H}  \phi_I + \lambda_e^{IJr} \phi_I \phi_J E^c_k+ 
\lambda_d^{Ipq} \phi_IQ_pD^c_q  + h_u^{ij} \bar{H} Q_i U^c_j \label{S}
\eeq
where $\mu_I = (\mu ,\epsilon_i)$, 
$\frac{1}{2}h_e^{jk} = \lambda_e^{0jk}$, $\lambda_1^{ijk} = 
\lambda_e^{ijk}, $  $\lambda_d^{0jk} = h_d^{jk}$, 
 and $\lambda_2^{ijk} = \lambda_d^{ijk}$.
This superpotential contains $R$ violation if 
different coupling constants, as vectors
and matrices in $\phi$ space, choose different directions 
to be the Higgs.  The differences between
the various directions chosen to be the
Higgs  parametrize the amount of $R$ violation
present, and are independent of basis. 
Therefore, to obtain
basis-independent measures of $R$ violation~\cite{DE},
one should first consider
a lepton Yukawa interaction, which defines
a plane spanned by a Higgs and lepton in $\phi$ space,
then use another interaction to fix one 
direction in that plane as the
Higgs, with the orthogonal direction
being defined as the lepton.  If some other
interaction chooses as the Higgs a direction
that has a non-zero
projection onto the direction previously chosen 
to be a lepton,
this yields a measure of $R$ violation that
is independent of the basis chosen
in $\phi$ space. Note that at least one Yukawa coupling
must enter into any such measure of $R$ violation,
because, in the absence of the Yukawa couplings, all the directions
in $\phi$ space are equally good choices to be the Higges, and 
it is not possible to define $R$ violation.

A number of algebraic expressions, which
correspond to sums of closed supergraphs, were computed in DE using
this algorithm:
\beq
\delta_1 = \frac{(\mu^{I*} \lambda_e^{IJp} \lambda_d^{Jqr*})
(\mu^{K} \lambda_e^{KLp*} \lambda_d^{Lqr})}
 {|\mu|^2 |\lambda_e|^2 |\lambda_d|^2} \label{sug1}
\eeq
\beq
\delta_2 = \frac{\mu^{I *} \lambda_e^{IJp} \lambda_e^{JKq *} 
  \lambda_e^{KLq} \lambda_e^{LMp *} \mu^M
  -1/2 ( \mu^{I *} \lambda_e^{IJp} \lambda_e^{JKq *} \mu^K )( \lambda_e^{LMp *}
  \lambda_e^{LMq} )}
{|\mu|^2 Tr[ \lambda_e^{p*} \lambda_e^p \lambda_e^{q*} 
\lambda_e^q]}  
 \label{sug2}
\eeq
\beq
\delta_3= \frac{\lambda_d^{Irs *} \lambda_e^{IJp} \lambda_e^{JKq *} 
  \lambda_e^{KLq} \lambda_e^{LMp *} \lambda_d^{Mrs}
  -1/2 ( \lambda_d^{Irs *} \lambda_e^{IJp} \lambda_e^{JKq *} \lambda_d^{Krs})
 (  \lambda_e^{LMp *} \lambda_e^{LMq} ) }
{|\lambda_d|^2 Tr[ \lambda_e^{p*} \lambda_e^p \lambda_e^{q*} 
\lambda_e^q]}  
 \label{sug3}
\eeq

\bea
\delta_4 = \frac{Tr 
[\lambda_e^p \lambda_e^{q*} \lambda_e^r \lambda_e^{r*} \lambda_e^{q} 
\lambda_e^{p*}]
+ Tr 
[\lambda_e^p \lambda_e^{q*} \lambda_e^r \lambda_e^{p*} \lambda_e^{q} 
\lambda_e^{r*}]}
{ Tr 
[\lambda_e^p \lambda_e^{p*} \lambda_e^r \lambda_e^{r*} \lambda_e^{q} 
\lambda_e^{q*}]
} ~~~~~~~~~~~~~~~~~~~~~~~~~~~\\ \nonumber
 -\frac{1}{2} \frac{  Tr 
[\lambda_e^p \lambda_e^{q*} \lambda_e^r \lambda_e^{p*}] Tr[ \lambda_e^{q} 
\lambda_e^{r*}]
 +  Tr 
[\lambda_e^r \lambda_e^{q*} \lambda_e^p \lambda_e^{r*}] Tr[ \lambda_e^{q} 
\lambda_e^{p*}]}
{ Tr 
[\lambda_e^p \lambda_e^{p*} \lambda_e^r \lambda_e^{r*} \lambda_e^{q} 
\lambda_e^{q*}]
} \label{sug4}
\eea

\beq
\delta_5 = \frac{(\lambda_d^{Ist*} \lambda_e^{IJp} \lambda_d^{Jqr*})
(\lambda^{Kst} \lambda_e^{KLp*} \lambda_d^{Lqr})
+ (\lambda_d^{Ist*} \lambda_e^{IJp} \lambda_d^{Jqr*})
(\lambda_d^{Kqt} \lambda_e^{KLp*} \lambda_d^{Lsr})}
 { |\lambda_e|^2 |\lambda_d|^4} \label{sug5} ~~~~.
\eeq
Here the traces are over the capitalized $\phi$ indices,
and the lower-case indices correspond to quark and
right-handed lepton generations, which are also summed~\footnote{Note 
that we have rewritten slightly $\delta_4$, as compared to
the expression in DE.  If one sums
over all the right-handed lepton indices, the expression above is
equivalent that in
to DE. However, if one does {\it not} perform
the right-handed index sum, the above expression vanishes
in the absence of $R$ violation, whereas the 
version in DE does not.}.

Some of the couplings in eqn (\ref{S}) yield interactions of particles
in the $\phi$ subspace with particles in other 
multi-dimensional spaces, {\it e.g.} $E^c, Q, D^c$. The
magnitudes of the couplings therefore
depend on the choice of basis in these other 
spaces. In DE, this was resolved by summing
over all the indices: those of $\phi$, of
the quarks, and of the right-handed leptons. 
This produced basis-independent expressions that
measured the total $L$ violation.
However, expressions that measure the
magnitude of $R$ violation for a given
quark or lepton generation would have additional phenomenological
applications. It would also be interesting to
know how many independent invariants there are:
it is clear that, if one {\it did not} do the quark and
right-handed lepton sums in the expressions
(\ref{sug1}) --(\ref{sug5}), there would
be more ``invariants'' than $R$-violating
couplings, so some of these invariants must be  redundant.

\section{Specification of the Superfield Bases}

There are various possible choices for
the $Q, D^c$ and $E^c$ bases. The most useful ones
are presumably those that diagonalise the propagators
in flavour space. In the absence of soft SUSY breaking,
this diagonalisation can be performed at the superfield level.
Moreover, we are neglecting the Higgs vacuum expectation
values, as is appropriate for conditions in the early
Universe. We may therefore choose the $Q$ basis such that
\beq
\sum_{Ip} \lambda_d^{Iqp}\lambda_d^{Irp*} + 
\sum_s h_u^{qs} h_u^{rs *} \propto \delta^{qr}  ~, 
\label{cdn1}
\eeq
the $D^c$ basis such that
\beq
\sum_{Ip} \lambda_d^{Ipq}\lambda_d^{Ipr*} \propto \delta^{qr}  ~,
\label{cdn2}
\eeq
and the  $E^c$ basis such that
\beq
\sum_{IJ} \lambda_e^{IJq}\lambda_e^{IJr*} \propto \delta^{qr}  ~.
\label{cdn3}
\eeq
These conditions ensure that the sums of the
vacuum polarization supergraphs are flavour
diagonal for each of the corresponding types of particle:$Q, D^c, E^c$.
For example, the conditions (\ref{cdn1}, \ref{cdn2}, \ref{cdn3})
ensure that the one-loop propagators are flavour-diagonal at high
temperatures in the early universe, when soft SUSY breaking and Higgs
vacuum expectation values can be neglected.
In the absence of $R$-parity breaking, the above prescription
yields the Standard Model mass-eigenstate basis for the
$E^c$ and $D^c$: writing $\lambda_e^{0jk} = \frac{1}{2} h_e^{jk}$
(the Yukawa coupling matrix of the leptons) and
$\lambda_e^{ijk} = 0$, equation (\ref{cdn3}) is
equivalent to  $h_e^{\dagger}  h_e$ being diagonal. 
This is the usual mass-eigenstate basis for
the singlet right-handed charged leptons in the Standard Model, and
a similar argument applies to $D^c$. 

\section{Counting the Flavour-Dependent and Basis-Independent Invariants
in Two Generations}

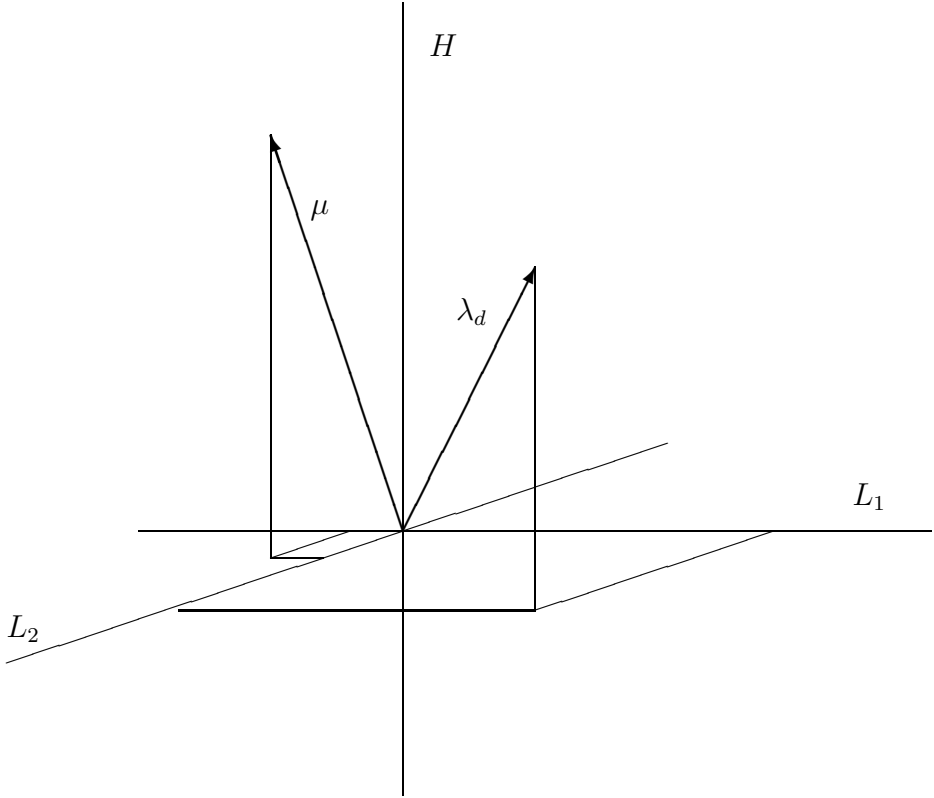
\begin{figure}
\begin{picture}(400,300)(-150,-100)
\thicklines
\put(0,0){\vector(1,2){50}}
\put(0,0){\vector(-1,3){50}}
\thinlines
\put(0,-100){\line(0,3){300}}
\put(0,0){\line(3,1){100}}
\put(0,0){\line(-3,-1){150}}
\put(-100,0){\line(3,0){300}}
\put(50,100){\line(0,-3){130}}
\put(50,-30){\line(3,1){90}}
\put(50,-30){\line(-3,0){135}}
\put(-50,150){\line(0,-3){160}}
\put(-50,-10){\line(3,0){20}}
\put(-50,-10){\line(3,1){30}}
\put(-35,120){$\mu$}
\put(20,80){$\lambda_d$}
\put(10,180){$H$}
\put(170,10){$L_1$}
\put(-150,-40){$L_2$}
\end{picture}
\caption{Directions chosen by the interactions
$\mu, \lambda_d^{pq}$ and $\lambda_e^i$ in  the
two-generation $\phi$ space: $i,p,q = 1,2$. 
The lepton Yukawa couplings $\lambda_e^1$ and $\lambda_e^2$
each choose a plane which is  spanned by the Higgs and a lepton. The
$\lambda_e$ choice of the Higgs is therefore the intersection
of the two planes, here labelled $\hat{H}$.
In general, $\mu$ and  the $\lambda_d^{pq}$ each
choose a preferred direction to identify as the Higgs: their
projections onto the choice of $L_1$ and $L_2$ made by
the $\lambda_e$ are basis-independent measures of
$R$-parity violation. Note that there are in principle four
distinct $\lambda_d^{pq}$: for simplicity, we show only one.}
\label{f2}
\end{figure}

Having chosen a well-motivated  basis
for the quarks and right-handed leptons,  we can now construct
invariants that measure the violation of individual lepton flavours,
where these are defined to be associated with
the right-handed leptons $E^c_i$. To illustrate how this may
be done, we first consider as a warm-up exercise a simplified
model with just two generations, whose geometry is easier to
visualise than the fully realistic three-generation case.
If one simply considers
all the expressions listed in equations (\ref{sug1}) - (\ref{sug5}),
and does not perform the quark and singlet-lepton sums,
one obtains 42 invariants in the two-generation model: eight from
$\delta_1$,
two from $\delta_2$, eight from $\delta_3$, none from $\delta_4$,
and 24 from $\delta_5$~\footnote{The corresponding number in the full 
three-generation model
would be 321 invariants.}. It is clear
that this is overcounting, because in
two generations there are only twelve
possible $R$-violating couplings: two from each of $\mu$
and the four $\lambda_d$, and one from each $\lambda_e$~\footnote{There
are 39 such couplings in the
three-generation model.}.
The reason that there are more invariants
than $R$-violating couplings is as follows.
The different couplings $\mu, \lambda_d, \lambda_e$, ...
choose different directions in $\phi$ space to be
the Higgs and the leptons, as seen in Fig.~\ref{f2}. 
The various invariants
measure the differences between these vectors,
which is a geometric invariant independent of the basis chosen in the
$\phi$ space. However, it is geometrically
clear that if one knows the vectors in $\phi$ space corresponding to 
the differences between 
$\mu$ and $\lambda_d^{11}$  and between 
$\mu$ and $\lambda_d^{12}$, then the invariant
measuring  the difference between 
$\lambda_d^{11}$  and $\lambda_d^{12}$ is not providing
any new information. 

Given the excess of invariants, we now analyze the number of
independent invariants,
and express the extra dependent ones in terms of
an independent basis set. This is simpler in two generations,
because  the $\phi$ space is three-dimensional,
and hence easier to visualize. We expect to find
10 independent invariants in two generations, because,
if one chooses to work in the basis where
$\mu = (\mu, 0 , 0)$, two $L$-violating 
couplings arise from each of the four $\lambda_d^{pq}$, and
one $L$-violating coupling from each of the $\lambda_e$.
This expectation of finding ten independent invariants is
indeed confirmed by our analysis.
 
The $\lambda_e$  have 
antisymmetric indices, so they have a zero eigenvector
in this two-dimensional model.
This means that they can be pictured as a plane in
$\phi$ space, spanned by the Higgs and a lepton:
\beq
\lambda_e^{IJk}  = |\lambda_e^k|
 (\hat{H}^I \hat{L}_k^J - \hat{L}_k^I \hat{H}^J)
\label{lame}
\eeq
where  $| \lambda_e^i | = 
 Tr [ \lambda_e^i \lambda_e^{i \dagger}]$.
The intersection of the two planes chooses
a direction  for the Higgs, which is here labelled  $\hat{H}$. The
directions orthogonal to $\hat{H}$ in the planes
defined by the $\lambda_e^i$
are labelled by $\hat{L_i}$ ($i = 1,2$). The lepton directions $\hat{L_i}$
are orthogonal by eqn (\ref{cdn3}):
\beq
 0 = \lambda_e^{IJ1} \lambda_e^{JI2} \propto 
(\hat{H}^I \hat{L}_1^J - \hat{L}_1^I \hat{H}^J)
(\hat{H}^J \hat{L}_2^I - \hat{L}_2^J \hat{H}^I) = 
-2  |\hat{H}|^2 \hat{L} _1 \cdot \hat{L}_2
\eeq
These definitions of $\hat{H}$ and $\hat{L}_i$
provide the axes in Fig.~\ref{f2}. 

There is $R$ violation
in this two-generation model if the direction chosen to
be the Higgs by the $\lambda_e$ couplings differs from
the direction chosen by $\mu$ or the $\lambda_d$ couplings. 
The lepton-number violation for
each flavour can be parametrized by taking the inner
product between $\mu$ or $\lambda_d$ -
directions in $\phi$ space that should
correspond to the Higgs - and $\hat{L_i}$ as defined
by the $\lambda_e$. This gives the following 10 independent
invariants
\beq
\delta_2^{12}   =   \frac{| \mu \cdot \hat{L}_1|^2 }{|\mu|^2} 
=  \frac{\mu^{I *} \lambda_e^{IJ1} \lambda_e^{JK2 *} 
  \lambda_e^{KL2} \lambda_e^{LM1 *} \mu^M 
 -1/2 ( \mu^{I *} \lambda_e^{IJ1} \lambda_e^{JK2 *} \mu^K )( \lambda_e^{LM1 *}
  \lambda_e^{ML2} )}
{|\mu|^2 Tr[ \lambda_e^{1*} \lambda_e^1 \lambda_e^{2*} 
\lambda_e^2]} 
 \label{sugg21}
\eeq
\beq
\delta_2^{21} =   \frac{| \mu \cdot \hat{L}_2|^2 }{|\mu|^2} 
 = \frac{\mu^{I *} \lambda_e^{IJ2} \lambda_e^{JK1 *} 
  \lambda_e^{KL1} \lambda_e^{LM2 *} \mu^M -1/2
   ( \mu^{I *} \lambda_e^{IJ2} \lambda_e^{JK1 *} \mu^K )( \lambda_e^{LM2 *}
  \lambda_e^{ML1} )}
{|\mu|^2 Tr[ \lambda_e^{1*} \lambda_e^1 \lambda_e^{2*} 
\lambda_e^2]}
 \label{sugg22}
\eeq
\beq
\delta_3^{12 rs} =   \frac{| \lambda_d^{pq} \cdot \hat{L}_1|^2 }{|\lambda_d^{pq}|^2} 
 =\frac{\lambda_d^{Irs *} \lambda_e^{IJ1} \lambda_e^{JK2 *} 
  \lambda_e^{KL2} \lambda_e^{LM1 *} \lambda_d^{Mrs}
 -1/2 ( \lambda_d^{Irs *} \lambda_e^{IJ1} \lambda_e^{JK2 *} \lambda_d^{Krs})
 (  \lambda_e^{LM1 *} \lambda_e^{ML2} ) }
{|\lambda_d^{rs}|^2 Tr[ \lambda_e^{1*} \lambda_e^1 \lambda_e^{2*} 
\lambda_e^2]}  
 \label{sugg31}
\eeq
\beq
\delta_3^{21 rs}  =   \frac{| \lambda_d^{pq} \cdot \hat{L}_2|^2 }
  {|\lambda_d^{pq}|^2} 
 = \frac{\lambda_d^{Irs *} \lambda_e^{IJ2} \lambda_e^{JK1 *} 
  \lambda_e^{KL1} \lambda_e^{LM2 *} \lambda_d^{Mrs}
 -1/2 ( \lambda_d^{Irs *} \lambda_e^{IJ2} \lambda_e^{JK1 *} \lambda_d^{Krs})
 (  \lambda_e^{LM2 *} \lambda_e^{ML1} ) }
{|\lambda_d^{rs}|^2 Tr[ \lambda_e^{1*} \lambda_e^1 \lambda_e^{2*} 
\lambda_e^2]}
 \label{sugg32}
\eeq
where the first expression in each case indicates the 
geometric interpretation of the invariant as indicated
in Fig.~\ref{f2}, and
the     $ \lambda_e \lambda_e$ 
traces in the second terms in all of these
formulae are zero in
the basis chosen by (\ref{cdn3}). 
The subscripts are derived from the
numbering of the $\delta$ invariants in equations (\ref{sug1}) to
(\ref{sug5}), and
the superscripts are the generation indices of
the right-handed leptons and quarks.
One can also construct other invariants,
corresponding to $ \delta_1$ and $\delta_5$ without their
generation sums. However, these can be expressed in terms
of the invariants  (\ref{sugg21}) - (\ref{sugg32}), 
as shown below, and
so are not independent. We therefore find 
the expected 10 invariant
measures of $R$ violation in this two-generation
model. We can express other possible invariants
in terms of the independent basis set (\ref{sugg21}) ---
(\ref{sugg32}). The combination of coupling constants
that appears in the relation between one set of
invariants and the other is  $\sqrt{\delta}$, which may be interpreted
as the length of a vector in $\phi$ space,
see  (\ref{sugg21}) --- (\ref{sugg32}).
For instance, the vector corresponding  to the difference
between $\lambda_d^{pq}$ and $\lambda_d^{rs}$,
is the vector sum of the differences between 
$\lambda_d^{pq}$ and $\mu$, and between $\mu$ and $\lambda_d^{rs}$.

In the basis chosen by the $\lambda_e$,
the invariants can written as:
\beq
 \sqrt{\delta_1^{i pq}} = 
\frac{(\lambda_d^{pq})_i \mu_0 - \mu_i (\lambda_d^{pq})_0}
{|\mu| |\lambda_d^{pq}|}
\label{ep1}
\eeq
\beq
 \sqrt{\delta_2^{ij }} =
  \frac{\mu_i}{ |\mu|} ~~~~, j \neq i
\label{ep2}
\eeq
\beq
  \sqrt{\delta_3^{ij pq}}  = \frac{(\lambda_d^{pq})_i}
{ |\lambda_d^{pq}|} ~~~~, j \neq i
\label{ep3} 
\eeq
\beq
 \sqrt{\delta_5^{pq i rs }}  = 
\frac{(\lambda_d^{pq})_i (\lambda_d^{rs})_0 - 
(\lambda_d^{rs})_i (\lambda_d^{pq})_0}{|\lambda_d^{pq}| |\lambda_d^{rs}|} ~~~~.
\label{ep4}
\eeq
It is clear from (\ref{ep1})
that one can express
\beq
\sqrt{\delta^{ipq}_1}  = \sqrt{\delta_3^{ij pq}( 1 - 
\delta_2^{ij} - \delta_2^{ji} )  } - \sqrt{\delta_2^{ij}
( 1 -\delta_3^{ij pq} - \delta_3^{ji pq})  } ~~~~ , ~j \neq i 
\eeq
and from (\ref{ep4}) one finds
\beq
\sqrt{\delta_5^{pq i rs}}  = \sqrt{\delta_3^{ij pq} ( 1 - 
\delta_3^{ij rs} - \delta_3^{ji rs})  } -\sqrt{ \delta_3^{ij rs}
( 1 -\delta_3^{ij pq} -\delta_3^{ji pq} )  } ~~~~ ,~ j \neq i
\eeq
where we have written
$\mu_0 = |\mu| [ 1 - \delta_2^{ij} - \delta_2^{ji}]^{1/2}$,
for instance.

\section{Extension to the Three-Generation Case}

The full three-generation model is
more complicated to analyze, both because it is more difficult
to visualize - the $\phi$ space is four-dimensional,
and because the $\lambda_e$ can contain $R$ violation
independently of the other interactions. We therefore
count invariants by making a particular basis choice that is
motivated by the coupling constants. The couplings $\mu$ and
$\lambda_d$ are still vectors in $\phi$ space
that would like to define the direction corresponding to
the Higgs,  but each  $\lambda_e$  is now geometrically
represented by {\it two} non-intersecting planes. To see this,
we recall that,
in three generations, any $\lambda_e$ (say $\lambda_e^1$) is a 
four-dimensional anti-symmetric matrix whose
eigenvalues can be labelled as $\pm i h_1$ and $\pm i c_1$. 
This means that there is a basis in which
\beq
\lambda_e^1 =  \left[
\begin{array}{cccc}
 0 & h_1 & 0 & 0 \\
-h_1 & 0 & 0 & 0 \\
0 & 0 & 0 & c_1 \\
0 & 0 & -c_1 & 0 
\end{array}
\right] ~~~~.
\label{twoplanes}
\eeq
Each of the two non-intersecting planes in $\phi$ space
represented by (\ref{twoplanes})
would like to be spanned by a Higgs and
a lepton. However, if a pair of the $\lambda_e^1$ eigenvalues
vanish, then $\lambda_e$ corresponds to a single plane,
as in the two-generation model discussed in the previous section,
and lepton flavour is conserved.
In the general non-degenerate case, both of these planes should be
spanned by a Higgs and a lepton field.
This means that
even a single $\lambda_e$ coupling does not in general conserve
lepton flavour, because  one right-handed lepton
can interact with two flavours of left-handed lepton.
Moreover, it would also like two directions in $\phi$ space to be
the Higgs.

In a similar fashion, a second coupling
$\lambda_e^2$ can also be pictured as two planes. The directions
in which the $\lambda_e^2$ planes intersect with
the $\lambda_e^1$ planes  can be chosen as the Higgs
and, for instance, $L_3$. The other leptons $L_1$ and $L_2$ can
be taken as the orthogonal directions
in the $\lambda_e^1$ planes. With this
choice, the Higgs and $L_3$ fields do not interact,
and nor do $L_1$ and $L_2$, so  $\lambda_e^2$ takes the form
\beq
\lambda_e^2 =  \left[
\begin{array}{cccc}
 0 & e_1 & e_2 & 0 \\
-e_1 & 0 & 0 & -b_2 \\
-e_2 & 0 & 0 & b_1\\
0 & b_2 & -b_1 & 0 
\end{array}
\right]
\eeq
The third coupling of this type, $\lambda_e^3$, is of generic form in the
basis
chosen in this way by $\lambda_e^1$ and $\lambda_e^2$. 
The counting of lepton-flavour-violating couplings therefore
yields one from $\lambda_e^1$, three from $\lambda_e^2$ and
five from $\lambda_e^3$.
However, there are also three conditions imposed on
these couplings by the choice of right-handed lepton
basis (\ref{cdn3}), so we expect 6 independent
invariant measures of $R$ violation. We could
take, for instance, $\delta_4^{iii}: i = 1, 2, 3$ and
$\delta_4^{ijk}, i \neq j \neq k$. This choice has the
feature that the $\delta_4^{iii}$ parametrize $L_i$ violation
in the $\lambda_e^i$, and the remaining three
invariants then measure the differences between
the bases chosen by the various $\lambda_e^i$.

As in the two-generation case, $\mu$ and the $\lambda_d^{pq}$
choose directions in $\phi$ space that should be identified
with the Higgs.  The components of these
vectors along the directions chosen by
the $\lambda_e$ to be leptons are invariant
measures of $R$ violation. We therefore need three
further invariants for each of $\mu$ and the $\lambda_d^{pq}$,
which can be taken to be:
\beq
\delta_2^{12} = \frac{\mu^{I *} \lambda_e^{IJ1} \lambda_e^{JK2 *} 
  \lambda_e^{KL2} \lambda_e^{LM1 *} \mu^M
   -1/2 ( \mu^{I *} \lambda_e^{IJ1} \lambda_e^{JK2 *} \mu^K )
  ( \lambda_e^{LM1 *} \lambda_e^{ML2} )}
{|\mu|^2 Tr[ \lambda_e^{1*} \lambda_e^1 \lambda_e^{2*} 
\lambda_e^2]}  
 \label{suggg212}
\eeq
\beq
\delta_2^{21} = \frac{\mu^{I *} \lambda_e^{IJ2} \lambda_e^{JK1 *} 
  \lambda_e^{KL1} \lambda_e^{LM2 *} \mu^M
  -1/2 ( \mu^{I *} \lambda_e^{IJ2} \lambda_e^{JK1 *} \mu^K )
( \lambda_e^{LM2 *}
  \lambda_e^{ML1} ) }
{|\mu|^2 Tr[ \lambda_e^{1*} \lambda_e^1 \lambda_e^{2*} 
\lambda_e^2]}  
 \label{suggg221}
\eeq
\beq
\delta_2^{31} = \frac{\mu^{I *} \lambda_e^{IJ3} \lambda_e^{JK1 *} 
  \lambda_e^{KL1} \lambda_e^{LM3 *} \mu^M
  -1/2 ( \mu^{I *} \lambda_e^{IJ3} \lambda_e^{JK1 *} \mu^K )
  ( \lambda_e^{LM3 *}
  \lambda_e^{ML1} ) }
{|\mu|^2 Tr[ \lambda_e^{1*} \lambda_e^1 \lambda_e^{3*} 
\lambda_e^3]}  
 \label{suggg231}
\eeq

\beq
\delta_3^{12 rs}= \frac{\lambda_d^{Irs *} \lambda_e^{IJ1} \lambda_e^{JK2 *} 
  \lambda_e^{KL2} \lambda_e^{LM1 *} \lambda_d^{Mrs}
 -1/2 ( \lambda_d^{Irs *} \lambda_e^{IJ1} \lambda_e^{JK2 *} \lambda_d^{Krs})
 (  \lambda_e^{LM1 *} \lambda_e^{ML2} ) }
{|\lambda_d^{rs}|^2 Tr[ \lambda_e^{1*} \lambda_e^1 \lambda_e^{2*} 
\lambda_e^2]}  
 \label{suggg312}
\eeq
\beq
\delta_3^{21 rs}= \frac{\lambda_d^{Irs *} \lambda_e^{IJ2} \lambda_e^{JK1 *} 
  \lambda_e^{KL1} \lambda_e^{LM2 *} \lambda_d^{Mrs}
  -1/2 ( \lambda_d^{Irs *} \lambda_e^{IJ2} \lambda_e^{JK1*} \lambda_d^{Krs})
 (  \lambda_e^{LM2 *} \lambda_e^{ML1} ) }
{|\lambda_d^{rs}|^2 Tr[ \lambda_e^{1*} \lambda_e^1 \lambda_e^{2*} 
\lambda_e^2]}  
 \label{suggg32}
\eeq
\beq
\delta_3^{31 rs}= \frac{\lambda_d^{Irs *} \lambda_e^{IJ3} \lambda_e^{JK1 *} 
  \lambda_e^{KL1} \lambda_e^{LM3 *} \lambda_d^{Mrs}
 -1/2 ( \lambda_d^{Irs *} \lambda_e^{IJ3} \lambda_e^{JK1 *} \lambda_d^{Krs})
 (  \lambda_e^{LM3 *} \lambda_e^{ML1} ) }
{|\lambda_d^{rs}|^2 Tr[ \lambda_e^{1*} \lambda_e^1 \lambda_e^{3*} 
\lambda_e^3]}  
 \label{suggg331}
\eeq
Note that $| \mu \lambda_e^1 \lambda_e^3|^2$ is not included
in the above list, although $| \mu \lambda_e^1 \lambda_e^2|^2$ is.
This is because the r\^ole of the expressions (\ref{suggg212}) -
(\ref{suggg331})  is to determine whether $\mu$ and/or $\lambda_d$
share the same definition of the Higgs with the $\lambda_e$, whereas
the differences among the $\lambda_e$ have already
been accounted for by the $\delta_4$ invariants. The
purpose of the second $\lambda_e$  factor in expressions (\ref{suggg212})
- (\ref{suggg331}) is only to choose
a direction in the plane of the first $\lambda_e$ to
be the Higgs. The orthogonal direction is then the lepton, and one
projects $\mu$ or $\lambda_d$ onto the direction of this lepton field.  
If the three $\lambda_e$ happen to conserve $R$ parity,
which occurs if they each represent a single plane, and the three
planes intersect in a common direction, which is then defined
unambiguously to be the Higgs,
then $ \mu \lambda_e^1 \lambda_e^3 \sim \mu_1 h_1 h_3$
and $ \mu \lambda_e^1 \lambda_e^2 \sim \mu_1 h_1 h_2$.
Thus one measures the same $R$-violating coupling, 
whether the second matrix is 
taken to be $\lambda_e^2$ or $\lambda_e^3$. 
We therefore have 6 + 30 = 36 invariants in the three-generation
case, as expected. 

Another way to count invariants in the 
three-generation case, which may seem more transparent, is to choose the
Higgs to be the direction
corresponding to $\mu$, and the leptons to be
$L_i = \mu \cdot \lambda_e^i$. The right-handed
lepton basis must then be chosen such that 
\beq
\mu^{\dagger} \lambda_e^i \lambda_e^{j \dagger}  \mu \propto \delta^{ij}
\eeq
so that the $L_i$ are orthogonal in $\phi$ space. 
The $\lambda_e$ are then of the form
\beq
\left[
\begin{array}{cccc}
 0 & h_1 & 0 & 0 \\
-h_1 & 0 & \lambda_e^{121} &  \lambda_e^{131} \\
0 & -\lambda_e^{121} & 0 &  \lambda_e^{231} \\
0 & -\lambda_e^{131} &   -\lambda_e^{231} & 0
\end{array}
\right]~~,~~
 \left[
\begin{array}{cccc}
 0 & 0 & h_2 & 0 \\
 0 & 0 & \lambda_e^{122} &  \lambda_e^{132} \\
-h_2 & -\lambda_e^{122} & 0 &  \lambda_e^{232} \\
0 & -\lambda_e^{132} &   -\lambda_e^{232} & 0
\end{array}
\right]~~,~~
 \left[
\begin{array}{cccc}
 0 & 0 & 0 & h_3 \\
 0 & 0 & \lambda_e^{123} &  \lambda_e^{133}\\
 0& -\lambda_e^{123} & 0 &  \lambda_e^{233} \\
-h_3& -\lambda_e^{133} &   -\lambda_e^{233} & 0
\end{array}
\right]~~.
\eeq
The $ 9 \times 3 = 27$ invariants
\beq
\delta_1^{\ell pq} = \frac{|\mu^{I *} \lambda_e^{IJ \ell } 
   \lambda_d^{Jpq *}|^2 }
{|\mu|^2  | \lambda_e^{\ell}|^2 | \lambda_d^{pq}|^2} 
 \label{suggg1}
\eeq
then measure $L_{\ell}$ violation in the $\lambda_d^{pq}$,
and the nine invariants
 \beq
\delta_2^{ jl} = \frac{\mu^{I *} \lambda_e^{IJj} \lambda_e^{JKl *} 
  \lambda_e^{KLl} \lambda_e^{LMj *} \mu^M
  - .5 \mu^I \lambda_e^{IJ j*} \lambda_e^{JK l} \mu^{K*} Tr[ \lambda_e^j 
  \lambda_e^{l *}] }
{|\mu|^2 Tr[ \lambda_e^{j*} \lambda_e^j \lambda_e^{l*} 
\lambda_e^l]}  
\label{suggg2}
\eeq
measure $L_j$ violation in $\lambda_e^l$: 
\beq
\begin{array}{ll}
L_1   {\!\!\!\!\! /} ~~{\rm ~ in~} \lambda_e^1: & 
 \delta_2^{11}  \sim  | \lambda_e^{321}|^2  \\
L_1 {\!\!\!\!\! /}~ ~ {\rm ~ in~} \lambda_e^2: & 
 \delta_2^{12}  \sim  | \lambda_e^{122}|^2 + | \lambda_e^{132}|^2 \\
L_1 {\!\!\!\!\! /} ~~ {\rm ~ in~} \lambda_e^3: & 
\delta_2^{13}  \sim   | \lambda_e^{123}|^2 + | \lambda_e^{133}|^2 \\
L_2 {\!\!\!\!\!  /} ~~ {\rm ~ in~} \lambda_e^1: & 
\delta_2^{21}  \sim  | \lambda_e^{211}|^2 + | \lambda_e^{231}|^2 \\
L_2  {\!\!\!\!\! /} ~~ {\rm ~ in~} \lambda_e^2: & 
\delta_2^{22}   \sim  | \lambda_e^{312}|^2  \\
L_2 {\!\!\!\!\! /} ~~ {\rm ~ in~} \lambda_e^3: & 
\delta_2^{23}  \sim  | \lambda_e^{213}|^2 + | \lambda_e^{233}|^2 \\
L_3 {\!\!\!\!\! /}~ ~ {\rm ~ in~} \lambda_e^1: & 
\delta_2^{31}  \sim  | \lambda_e^{321}|^2 + | \lambda_e^{311}|^2 \\
L_3 {\!\!\!\!\! /} ~~ {\rm ~ in~} \lambda_e^2: & 
\delta_2^{32}  \sim | \lambda_e^{312}|^2 + | \lambda_e^{322}|^2 \\
L_3   {\!\!\!\!\! /} ~~{\rm ~ in~} \lambda_e^3: & 
\delta_2^{33}  \sim  | \lambda_e^{213}|^2 
\end{array}
\eeq
completing our second enumeration of a basis of independent
flavour-dependent measures of $R$ violation~\footnote{Note 
that, in the two-generation case, $\lambda_e^j$ can not violate $L$
and $L_j$ simultaneously. This is because $\lambda_e^2$
couples $E^c_2$ to either
$H L_1$, $H L_2$ or $L_1 L_2$. Therefore, in the two-generation
case it is easy to check that $\delta_2^{jj} = 0$.
However, in the three-generation case, $\lambda_e^{ijk}$
can violate $L$ and $L_k$ simultaneously via
the couplings $\lambda_e^{ijk}~ i,j, \neq k$,
and  this is measured by $\delta_2^{jj}$.}.

\section{Application to Lepton Flavour Violation in Early Cosmology}

In the hot plasma of the early Universe at temperatures
above the electroweak phase transition (EPT): $T \gappeq T_c \simeq 100$
GeV,
non-perturbative processes that violate $B + L$ \cite{BL} are in thermal
equilibrium. If $R$-parity non-conserving interactions
that violate all the $L_i$ are also in thermal equilibrium
above the EPT, then
any asymmetry in  $B$ or the $L_i$ gets washed out. 
This means that either the baryon asymmetry of the
Universe (BAU) observed today
was not present in the thermal plasma above the electroweak
phase transition, motivating scenarios in which it was made at
the EPT, 
or the $R$-violating couplings
are sufficiently small to be out of equilibrium above the EPT.
This argument has been used to place strong
constraints on $R$-violating Yukawa couplings \cite{CDEO,DR} of
order 
\beq
 \lambda_{R} \lappeq 10^{-7}
\label{CDEO}
\eeq
where $\lambda_R$ is some generic $R$-violating Yukawa
coupling. In this section we review these constraints in the
basis-independent approach of DE, with the added ingredient
that we discuss carefully the bounds on each lepton flavour
separately, using the flavour-dependent invariants isolated
in the previous sections~\footnote{For a more extensive 
discussion in the mass-eigenstate basis, see \cite{DR}.}.

We first expand
on the previous paragraph's discussion of
the origin of the bounds.
The minimal supersymmetric extension of the 
Standard Model (MSSM) has four global
quantum numbers: $B, L_1, L_2, L_3$ that are
conserved at the perturbative level, in the absence of
the $R$-violating couplings (\ref{S1}). However, 
anomalous non-perturbative processes  that
are in thermal equilibrium
above the electroweak phase transition, 
but irrelevantly weak today, violate the combination
$B+L$, so the global conserved
quantum numbers in the MSSM
are $B/3 - L_i$: $i: 1, 2, 3$. If one then introduces
the $R$-violating interactions of (\ref{S1}), there
are no conserved global quantum numbers left. This means
that, with the MSSM particle content, all asymmetries
in the plasma are washed out before the electroweak phase
transition if the $R$-violating interactions
are in thermal equilibrium. As has already been
mentioned, one possible origin for the
baryon asymmetry observed in the Universe
today, is to make the asymmetry at the transition,
which may be possible for certain areas
of SUSY parameter space \cite{Kimmo}. 
One can also create the asymmetry
from particles decaying after the EPT
\cite{latedecay,KRS}, or store it in particles
not in thermal equilibrium with the 
MSSM while the $R$-parity
violating interactions are present \cite{DH}.
For the rest of this paper, we will neglect
these scenarios, and assume that the BAU
was present in the plasma above the EPT,
and must be protected from the depredations
of the $R$-violating interactions.

We assume that non-perturbative $B+L$ violation
is in thermal equilibrium before the EPT,
and consider only the renormalisable
$R$-violating interactions of equation 
(\ref{S1})~\footnote{Non-renormalisable interactions 
are different \cite{CDEO3}, in that
they may be out of equilibrium before all the Standard Model interactions
come into equilibrium, in which case  the asymmetries
are not neccessarily washed out.}.
A sufficient condition for
the BAU to survive is that all the $R$-parity
violating interactions should be out of thermal
equilibrium before the EPT, which gives
the bound (\ref{CDEO}). However, this is
too restrictive, because it means that the asymmetries in all
three of the $B/3 - L_i$ are preserved, whereas
keeping one of them would be sufficient to preserve the BAU.

More precise bounds can be obtained by using
our invariants in an analysis based on chemical 
potentials \footnote{We follow her the naive kinetic
theory discussion of~\cite{DR,HT}, which is adequate
for our purpose. For a more complete analysis using
the grand canonical ensemble,
see~\cite{KS}.}.
We assume an interaction to be in 
effective chemical equilibrium in
the early Universe if the interaction rate exceeds
the expansion rate. If this condition is satisfied, 
the sum of the chemical potentials of the particles
participating in the interaction is zero \cite{LL}~\footnote{We 
work in a mass-eigenstate basis.}.
In the Standard Model and the MSSM, all the interactions
are in equilibrium once the temperature has dropped
to $\sim 1 - 10$ TeV, imposing~\cite{HT}
a set of linear equations on the chemical potentials
that we now review. Since the gauge bosons
have zero chemical potential, all the members
of a gauge multiplet share the same chemical potential.
The Higgs interactions in the Standard Model imply:
\bea
 q^i + h - u_R^j  = 0 \label{ce1}\\
q^i - h - d_R^j = 0\label{ce2} \\
\ell^i - h - e_R^i = 0  \label{ce3}
\eea
where the chemical potential of a particle is
labelled by the particle's name: $q$ is the left-handed
quark doublet, $h$ the Higgs doublet, and so on. The
flavour non-diagonal quark couplings to
the Higgs imply that all the left-handed quarks
have the same chemical potential, so $q^i = q^j \equiv q$,
and similarly for the different flavours of $u_R$ and $d_R$ quarks.  The 
anomalous $B+L$ violation  transforms left-handed quarks
into left-handed anti-leptons:
\beq
 3 N_g q + \sum_i \ell^i = 0
\label{sph}
\eeq
where $N_g$ is the number of quark generations. We therefore
have six linear equations for 10 unknowns.  The
four free parameters correspond to the
four conserved charges in the plasma: $Q_{em}$ and
the $B/3 - L_i$. The plasma can contain asymmetries associated
with the three conserved global quantum numbers
$B/3 - L_i$, but is required not to have an electric
charge~\footnote{Or hypercharge - it amounts to
the same constraint~\cite{HT}.} asymmetry. The electric
charge  density can be written as
\beq
n_{em} = \sum_a Q_a ( n_a - n _{\bar{a}})
\eeq
where the sum over $a$ runs over all the particles
in the plasma, $Q_a$ is the particle electric charge,
and $n_a (n_{\bar{a}})$ is the (anti-)particle number
density.  For masses
and chemical potentials much smaller than
the temperature, the particle asymmetry is : 
\beq
n - \bar{n} \simeq \frac{ g \mu T^2}{3} \left[ 1 + {\cal{O}}\left(
\frac{m^2}{T^2} \right) \right]  ~~~{\rm for~ bosons},
\label{bosonasym}
\eeq
and
\beq
n - \bar{n} \simeq \frac{ g \mu T^2}{6} \left[ 1 + {\cal{O}} \left(
\frac{m^2}{T^2} \right)\right] ~~~{\rm for~fermions},
\label{fermionasym}
\eeq
where $g$ is the number of internal degrees
of freedom of the particle, and
$m$ is the zero-temperature mass~\footnote{See
\cite{KRS,DR,leptonmass} for an  analysis including
these mass effects, and~\cite{KS} for an careful discussion
of the relation of the chemical potentials and masses to the asymmetries.
We use here $\mu$ as a generic chemical potential,
and hope that the distinction between the chemical
potential and the $\mu$ term in the MSSM superpotential
will be clear from the context.}.
If we neglect the mass contributions in (\ref{fermionasym})
and (\ref{bosonasym}), since their contribution
to the asymmetry is reduced with respect to
the first terms by $m^2/T^2 \ll 1$, the electric charge
density in the Standard Model is
\beq
n_{em} = N_g ( q + 2 u_R - d_R) - \sum_i \ell^i - \sum_i e_R^i + 2h = 0 
\label{Qmu}
\eeq
(This equation must be multiplied by three in
the MSSM at $T > m_{SUSY}$.)
In the MSSM, there are more than twice as many new
particles, and therefore many new chemical potentials,
but these turn out all to be equal to those
of the  Standard Model particles.
The gauginos are majorana particles, so can not store
any asymmetry and must have zero chemical potential. 
The gaugino-scalar-fermion interactions then imply
that the chemical potentials of the spartners ($\equiv \phi$) are equal
to those of the Standard Model particles($\equiv \psi$):
\beq
\phi = \psi
\label{gaugino}
\eeq
The only remaining new chemical potential 
is that of the second Higgs $\bar{H}$, which is
required by the mass term $\mu H \bar{H}$
to have equal and opposite chemical potential
to the Higgs $H$.

The solution of the equations of
chemical equilibrium in the absence of $R$
violation is easily obtained. One can represent
the chemical potentials in the MSSM as
a vector  in a ten-dimensional
space: $\vec{\mu} \equiv 
( \ell^1, \ell^2, \ell^3, e_R^1, e_R^2, e_R^3,
h, q, u_R, d_R$). The equilibrium conditions
(\ref{ce1} - \ref{sph}) and  (\ref{Qmu})  force this vector to be orthogonal
to the directions associated with  $Q_{em}$  and
the interactions: for
instance, the interaction (\ref{ce1})corresponds
to the direction (0,0,0,0,0,0,1,1,-1,0). The conservation
of the three $B/3 - L_i$ implies that the projections
of the chemical potential vector $\vec{\mu}$  onto  the
directions corresponding to $B/3 - L_i$ are constant,
fixing $\vec{\mu}$. The direction corresponding to
$B/3 - L_1$ is
\beq
\vec{v}_{\frac{B}{3} - L_1} \equiv
(-2,0,0,-1,0,0,2,1,1)
\eeq
and there are
similar expressions for  $\vec{v}_{\frac{B}{3} - L_2}$ and
$\vec{v}_{\frac{B}{3} - L_3}$.
it is important to note that these vectors  are {\it not} orthogonal to
the
interaction vectors (\ref{ce1} - \ref{sph}), nor to
the electric charge direction (\ref{Qmu}). This is to be
expected, since the Yukawa interactions can change the
ratio of left- and right-handed quarks,
although they cannot change the total quark number.
One cannot simply express $\vec{\mu}$ in terms
of the $\vec{v}_{\frac{B}{3} - L_i}$. Instead, one can subtract
from the $\vec{v}_{\frac{B}{3} - L_i}$ the components along the
directions corresponding to (\ref{ce1} - \ref{sph}) and
(\ref{Qmu}), leaving three vectors $\vec{V}_i$ that 
are the projections of the $\vec{v}_{\frac{B}{3} - L_i}$
onto the space allowed for $\vec{\mu}$. Then $\vec{\mu}$ 
can be expressed in terms of the $\vec{V}_i$, as:
\beq
\vec{\mu} = \eta_{\frac{B}{3} - L_i} \vec{V}_i
\eeq
where $\eta_{\frac{B}{3} - L_i}$ is the conserved ratio of 
$B/3 - L_i$ to photons. The baryon-to-photon ratio is therefore
\cite{KS}
\beq
\eta_B = \vec{\mu} \cdot \vec{B} = 
\sum_i \eta_{\frac{B}{3} - L_i} \vec{V}_i \cdot \vec{B}
= \frac{28}{79} \sum_i \eta_{\frac{B}{3} - L_i} 
\eeq
where $\vec{B} = (0,0,0,0,0,0,0,6,3,3)$.
 
Now suppose we include the $R$-violating interaction
$\lambda_3^{ijk} U^c_i D^c_j D^c_k$ of equation
(\ref{S1}) to the thermal
soup above the electroweak phase transition.  We assume for the moment
that all the $L_i$-violating couplings are zero or
small enough to satisfy (\ref{CDEO}), and 
neglect them. If $\lambda_3$  is in
thermal equilibrium, i.e., at least one $\lambda_{ijk} > 10^{-7}$
in the thermal mass-eigenstate basis, then there is
an additional constraint on the chemical potentials
\beq
u_R + d_R + d_R = 0 ~~~~.
\label{lambda3}
\eeq
In conjunction with (\ref{ce1} - \ref{sph}) and (\ref{Qmu}),
this constraint forces $q = 0$, so there is no baryon
asymmetry. An alternative way to formulate this
is to note
that the direction in chemical potential space corresponding
to $\vec{B}$ can be written as a linear combination of
the directions (\ref{ce1} - \ref{ce3}), (\ref{lambda3})  and (\ref{Qmu}),
to which $\vec{\mu}$ must be orthogonal. 
So, in the presence of the non-perturbative $B+L$
violation and even one of the $\lambda_3^{ijk}$,
the baryon asymmetry will be washed out, up to
lepton mass effects \cite{KRS,leptonmass}~\footnote{This caveat applies
when there are different asymmetries in the various lepton flavours,
but is unlikely to preserve a large enough asymmetry, since it would be
proportional to $(m_{\tau}/T)^2 $, see equations (\ref{bosonasym})
and (\ref{fermionasym}).
A similar argument  can be made using the slepton
mass differences in the MSSM \cite{DR},
which could preserve a larger asymmetry if
their mass differences are large and
the sleptons are light enough to be
still present in the thermal bath
when the  sphalerons go out of equilibrium.}. The
baryon-number-violating couplings $\lambda_3^{ijk}$
should therefore satisfy (\ref{CDEO}). This is unsurprising,
since any one of the $\lambda_3$ interactions would violate 
all the $B/3 - L_i$, at least
one of which  needs to be conserved~\footnote{Note
that the $\lambda_3^{ijk}$ alone, in the
absence of $B+L$ violation, is not sufficient to
take the BAU to zero in the presence of
a lepton asymmetry \cite{DR}. This is because 
the lepton asymmetry carries electric charge,
which must be cancelled by an asymmetry in the
higgs and quarks. In the presence of
$\lambda_3$, the  electric charge
asymmetry in the quarks  also carries baryon number.}.

We now consider the $L_i$-violating couplings
present in (\ref{S1}), assuming that the $\lambda_3^{ijk}$
satisfy (\ref{CDEO}). We wish to find the minimal bounds
on the various lepton-flavour-violating rates
that ensure that an asymmetry in at least one of the $B/3 - L_i$
is preserved. If we include interactions that violate all of
$L_1, L_2$ and $L_3$, they imply 
the chemical potential relations
\beq
h = \ell^i
\label{above}
\eeq
for $i = 1, 2, 3$, implying that all the  asymmetries are washed out.
Even the lepton mass effects cannot help in this case, because
all the lepton flavours would have the same asymmetry by 
(\ref{above}). However, if the interactions in equilibrium
conserve a lepton flavour, say $L_1 \equiv L_e$, then the asymmetry
in $B/3 - L_1$ is conserved, and a baryon asymmetry
\beq
\eta_B =  \frac{28}{79} \eta_{\frac{B}{3} - L_1}
\eeq
will remain. Therefore, to preserve the BAU, the interactions violating
B and at least one of the
$L_i$ must be out of equilibrium. In our basis-independent
formulation, requiring that the
asymmetry in $B/3 - L_1$ be conserved  will translate into a bound
on  the invariants $\delta$ involving the
Yukawa coupling $\lambda_e^1$.

To set a bound on the $L_e$-violating couplings
or invariants, 
we need an estimate of the $L_e$-violating rates in
the early Universe at $T \gsim 100$ GeV. 
If one worked in a random basis, one would identify $\mu^1$, 
$(\lambda_d^{pq})^1$ -
the $\phi^1$ element of the $\mu$ and $\lambda_d^{pq}$ vectors,
$(\lambda_e^1)^{1j}: j: 1, 2, 3$ and  $(\lambda_e^2)^{1J},
 (\lambda_e^3)^{1J} , J:0, ..., 3$  as  $L_e$-violating
couplings, and require that the
rates for the processes mediated by these couplings to be out
of equilibrium.
The rate associated with any $R$-violating
Yukawa coupling $\lambda_R$  can be estimated as
\beq
\Gamma_{yuk} \simeq 10^{-2} \lambda_R^2 T ~~.
\eeq
and the rate
for the mass corrections to gauge interaction
rates can be taken as   
\beq
\Gamma_{mass} \simeq 10^{-1} \frac{m^2}{T^2} g^2 T
\eeq
where the inverse powers of ten account for factors
of $4 \pi$, etc..
Requiring that all the ``$L_e$-violating"  rates listed above
be less than the expansion rate at $T \sim 100$ GeV would give
the bounds
\beq
\lambda_R < 10^{-7}
\label{yuk}
\eeq
\beq
 \mu^1 < 100 ~ {\rm keV}
\label{mass}
\eeq
However, in the absence of a careful discussion of the basis
dependence of this analysis, it
is unclear to which elements of the coupling
constant vectors and matrices (\ref{yuk}) applies, because 
it is not obvious
which direction in $\phi$ space is $L_1$, and (\ref{mass})
is wrong because the gauge interactions are diagonal
in any basis for $\phi$ space, and so cannot violate
$R$ parity. The correct bounds can be calculated by 
working in the $T = 100$ GeV
mass-eigenstate basis, 
or by estimating the rates in some basis-independent
way. The drawback of  the thermal mass-eigenstate basis
is that it is not the same as the zero-temperature 
Standard Model mass-eigenstate basis we live in today.  It would
is therefore desirable to find some basis-independent  way
of estimating rates associated with the invariants that
parametize $L_e$ violation. 

We first consider which invariants must be zero 
for $L_e$ to be conserved.  We would like to find  a direction
in $\phi$ space that can be chosen as the left-handed lepton
associated with $E^c_1$. For this to be possible, $\lambda_e^1$ 
must correspond to only one plane in $\phi$ space, spanned by 
$E_L$ and the Higgs, and the other two $\lambda_e$ can
only intersect that plane in one direction, which is then
the Higgs. This then means that these other two
$\lambda_e$ can also only consist of one plane
each, because if either of these consisted of two orthogonal planes,
they would intersect the $\lambda_e$ plane twice. We conclude that
the three $\lambda_e^i$ must each consist of a single
plane, and that $\lambda_e^1$ intersects at least one of the $\lambda_e^2,
\lambda_e^3$ planes, which chooses the direction
for the Higgs: $L_1$ is then the perpendicular 
direction in the $\lambda_e^1$ plane. If $\lambda_e^1$ intersects both
$\lambda_e^2$ and $\lambda_e^3$, it must do so in the
same direction~\footnote{Note that $R$ violation in
the $\lambda_e$ is still possible: e.g., $\lambda_e^1$
could choose the plane (0-1), $\lambda_e^2$ the
plane (0-2) and $\lambda_e^3$ the
plane (2-3). In this case, the invariants  $\delta_4^{123}, \delta_4^{132}$
and $\delta_4^{ijk}$
with $i,j,k = 2,3$ could still be non-zero.}.
The $\lambda_d$ and $\mu$ must have no components in
the direction $E_L$, to ensure that  $L_e$  is conserved.

This geometry must now be
translated into statements about the vanishing of certain
invariants. First, we need  $\delta_1^{1 pq} \sim 
|\mu \lambda_e^1 \lambda_d^{pq}|^2 = 0$ for all $p,q$. Recall
that $\lambda_e^1$ represents a single plane,
as in the two generation case, so is
of the form (\ref{lame}). Setting this invariant
to zero 
means that $\mu$ and $\lambda_d^{pq}$ do not have
components along a direction in the $\lambda_e^1$ plane.
Secondly, for similar reasons
we also need $\delta_5^{pq1rs} = 0, \forall~ p,q,r,s$, though this
invariant is not independent, as previously discussed.
Thirdly, we need $\delta_2^{1j} = 0$ for all $j$, to ensure that the
direction in the $\lambda_e^1$ plane chosen to be the Higgs by
$\mu$ is the same as the direction chosen by the $\lambda_e^j$, and there
is again a similar  argument for $\delta_3^{1pq}$.
Finally, we need  $\delta_4^{i1j} = 0$. This implies that
there is no $L_1$ violation among the $\lambda_e$:
as previously argued, each $\lambda_e$
must represent a single plane, and if both
$\lambda_e^2$ and $\lambda_e^3$ intersect the
$\lambda_e^1$ plane, it must be in the same direction.
Writing $\lambda_e^1$ in the form (\ref{lame}),
where  $\hat{H}$ is the  direction in
which the three $\lambda_e$ intersect, 
it is clear that $\lambda_e^2 \lambda_e^1 \lambda_e^3 = 0$.
The $\delta_4^{11j}$ must also be zero, because
the last two terms cancel the first two in the
expression (\ref{sug4}), in the absence of $L_1$ violation.

We now know which invariants must be zero to
conserve $L_e$ exactly. However, this is not
necessary to preserve the baryon asymmetry:
we only need $L_e$ violation to be out of
thermal equilibrium at $T \sim 100$ GeV. We 
therefore need to be able to estimate a
rate for each invariant. It is possible to visualise
any given invariant (e.g., $\delta_1^{1 pq}$), made
up of three coupling constants ($\mu, \lambda_e^1, \lambda_d^{pq}$ in
this case) as representing  the squared amplitude for the process
mediated by the ``middle" interaction ($\lambda_e^1$ in this case),
with thermal masses due to the ``outside
interactions" ($\mu, \lambda_d$ in this case) on the external $\phi$ legs
in one of the amplitudes: see figure 2. For
the invariant to represent a process taking place
in the early Universe, the interactions appearing in
it would have to be in thermal equilibrium. In the case of
$\delta_1^{1pq}$, this would require $|\mu|$ to be in equilibrium,
so that it gives a mass to a direction in $\phi$ space
that could be the Higgs. Also, the row of the matrix
$\lambda_e^1$  that transforms this ``Higgs"
into a lepton would have to be in equilibrium,
and we would need the component
of $\lambda_d^{pq}$ coupling to this
lepton direction to be in equilibrium. We can estimate self-consistently
the coupling constant for the $\lambda_e^1$-mediated process 
to be $|\lambda_e^1|$, as will be explained below, and
the relevant component squared
of $\lambda_d^{pq}$ has magnitude $\delta_1^{1 pq} \times
| \lambda_d^{pq}|^2$. We  could just as well
have made this argument starting with
$\lambda_d^{pq}$, in which case the relevant $R$-violating 
coupling squared would
be the mass $\delta_1^{1pq} \times |\mu |^2$, so the basis-independent 
bound on $L_e$ violation associated with  the
invariant $ \delta_1^{1pq}$ is 
\beq
\delta_1^{1pq} \times \min \{ \Gamma_{|\lambda_d^{pq}|}, \Gamma_{|\mu|} \} < H
\eeq
The rate associated with $|\lambda_e^1|$ must also be in
equilibrium, because otherwise no direction in
$\phi$ space would be defined as $E_L$.

Consider now the invariants $\delta_2^{1j}$.
These can also be imagined as corresponding to Feynman
diagrams  with thermal masses  that should belong to
the Higgs on both $\phi$ lines, and a Yukawa interaction
that converts a Higgs into $E_L$ internally. Since one
of the $\phi$ lines is supposed to be $E_L$, one of
the thermal  masses that should belong to
a Higgs but is on the $E_L$ line must be out of equilibrium.  
So we obtain the similar bound
\beq
\delta_2^{1j} \times \min \{ \Gamma_{|\lambda_e^{j}|}, \Gamma_{|\mu|} \} < H
\eeq
provided that $|\lambda_e^1|$ is in thermal equilibrium~\footnote{It 
may be noted
that the $\lambda_e$ give thermal masses to both the Higgs and
the leptons, so that the $\phi$ leg with the thermal mass induced
by $\lambda_e$
does not have to be a Higgs, but
could just as well be a lepton. This is true,
but in the invariant $\delta_2^{11}$ this contribution
is cancelled by the second term in the invariant. So 
$\delta_2^{11} \times \Gamma_{|\lambda_e^1|} < H$
means that the $R$-violating contribution
is out of equilibrium.}.

Our previous assertion that one
can estimate the coupling constant for Higgs-lepton conversion 
due to $\lambda_e$ as  $|\lambda_e^1|$ merits some
further discussion. To see
why this statement might be suspect, suppose that the Higgs
is $\phi_0$ and $E_L$ is $\phi_1$, so that the 
coupling constant for the $H E_L E^c$ interaction
is $(\lambda_e^1)^{01}$. If $(\lambda_e^1)^{23}$
was much larger, it would make a greater contribution 
to $|\lambda_e^1|$, and one could  overestimate
the rate for the  $H E_L E^c$ process. However,
this does not happen, because any element
of the matrix $\lambda_e^1$ that does not
mediate   the $H E_L E^c$ process is an $L_1$-violating coupling,
and is therefore required to be smaller than
the $L_1$-conserving one.

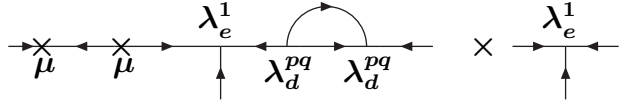
\begin{figure}
\begin{boldmath}
\begin{center}
\begin{picture}(250,100)(0,0)
\ArrowLine(0,20)(15,20)
\ArrowLine(40,20)(15,20)
\ArrowLine(40,20)(80,20)
\ArrowLine(110,20)(80,20)
\ArrowLine(110,20)(140,20)
\ArrowLine(160,20)(140,20)
\ArrowLine(80,0)(80,20)
\ArrowArcn(120,20)(15,180,0)
\ArrowLine(190,20)(210,20)
\ArrowLine(230,20)(210,20)
\ArrowLine(210,0)(210,20)
\Text(185,20)[r]{$\times$}
\Text(85,30)[r]{$\lambda_e^1$}
\Text(215,30)[r]{$\lambda_e^1$}
\Text(18,12)[r]{$\mu$}
\Text(48,12)[r]{$\mu$}
\Text(18,20)[r]{$\times$}
\Text(48,20)[r]{$\times$}
\Text(115,10)[r]{$\lambda_d^{pq}$}
\Text(145,10)[r]{$\lambda_d^{pq}$}
\end{picture}
\end{center}
\end{boldmath}
\label{F1}
\caption{ Representation of the invariant $\delta_1^{1pq}  \propto
|\mu^{\dagger} \lambda_e^1 \lambda_d^{pq *}|^2$  as the Feynman
supergraph for a scattering process mediated by $\lambda_e^1$, with
thermal masses on the external legs induced by the interactions
$\mu$ and $\lambda_d^{pq}$.} 
\end{figure}

We conclude that in basis-independent notation, the 
appropriate cosmological bounds are
\beq
\delta_1^{1pq} \times \min \{ \Gamma_{|\mu|}, \Gamma_{| \lambda_d^{pq}|} \} < H
\eeq
\beq
\delta_2^{1j} \times \min \{ \Gamma_{ |\mu| }, \Gamma_{ | \lambda_e^j | } \} 
< H 
\eeq
\beq
\delta_3^{1jpq} \times \min \{ \Gamma_{|\lambda_e^j|}, 
\Gamma_{| \lambda_d^{pq}|} \} < H  
\eeq
\beq
\delta_4^{j1k} \times \min \{ \Gamma_{|\lambda_e^j|}, 
\Gamma_{|\lambda_e^k|} \} < H  
\eeq
\beq
\delta_5^{rs1pq} \times 
 \min \{ \Gamma_{|\lambda_d^{rs}|}, \Gamma_{| \lambda_d^{pq}|} \} < H
\eeq
where the rates $\Gamma$ and $H$ are to
be evaluated at $T \sim T_c \sim 100$ GeV,
using the rate estimates
(\ref{yuk}) and (\ref{mass}). This provides the following bounds on
the invariants:
\beq
\delta_1^{1pq} |\lambda_d^{pq}|^2 <10^{-14}  ,~~p,q:1..3  \label{65}
\eeq
\beq
{\delta_2^{1j}}{|\lambda_e^j|^2} < 10^{-14}  ,~~j:1..3 
\eeq
\beq
{\delta_3^{1jpq}} \times {\min \{ |\lambda_e^j|^2, |\lambda_d^{pq}|^2 \}} 
 < 10^{-14} ,~~j,p,q: 1..3 
\eeq
\beq
{\delta_4^{k1j}} \times {\min \{ |\lambda_e^j|^2, |\lambda_e^{k}|^2 \}}
  < 10^{-14}, ~~ j,k:1..3 
\eeq
\beq
{\delta_5^{rs1pq}} \times {\min \{ |\lambda_d^{rs}|^2, |\lambda_d^{pq}|^2 \}} 
 < 10^{-14} ,~~r,s,p,q: 1..3  \label{69}
\eeq
where $i,j,k$ are lepton family indices, and $p,q,r,s$ are
quark family indices. We can conservatively 
estimate the magnitudes of the various coupling constant
vectors and  matrices as taking their Standard Model values.
If the couplings satisfy these
bounds, any asymmetry in $B/3 - L_1$ 
present in thermal equilibrium in
the plasma before the EPT will be preserved. Of course,
it could just as well be an asymmetry
in some other lepton flavour that is conserved,
in which case one would obtain identical bounds, but with
the lepton-flavour index ``1'' replaced by ``2''
or ``3''.

The results (\ref{65}) --- (\ref{69}) 
can be translated into basis-dependent
bounds on various individual coupling constants,
that may be more intuitive.
For instance, to preserve an asymmetry in $B/3 - L_1$,
the $R$-violating coupling $\mu^1$ should satisfy
\beq
\frac{\mu^1}{|\mu|}h_b < 10^{-7}
\eeq
in the thermal mass-eigenstate basis,
where $h_b$ is the bottom Yukawa coupling.
This can be seen by writing out the invariant
$\delta_1^{133}$ in the thermal mass-eigenstate 
basis, and requiring that each term in the sum
making up $\delta_1^{133}$ satisfy the bound (\ref{65}).
Similarly,
the other $L_e$-violating couplings should
satisfy
\beq
(\lambda_d^{pq})^1 < 10^{-7}  
\eeq
\beq
(\lambda_e^1)^{1j}, (\lambda_e^1)^{j1} < 10^{-7} ~~j:1..3
\eeq
\beq
(\lambda_e^{2,3})^{1J}, (\lambda_e^{2,3})^{J1} < 10^{-7} ~~J:0..3
\eeq
in the thermal mass-eigenstate basis.

\section{Conclusions}

In the Standard Model, there are no renormalisable
lepton-number-violating operators, 
and lepton flavour number is conserved because
the neutrinos are massless. However, with the
MSSM particle content, numerous renormalisable lepton-number-violating
operators can be constructed. This is because the leptons
have the same gauge quantum numbers as a Higgs field,
so the sleptons can interact with the same particles 
as the Higgs. If any of these lepton-number non-conserving operators
are present in the Lagrangian, there is no longer a global symmetry
defining lepton number, so a ``lepton"
and a ``lepton-number-violating interaction"
are no longer uniquely defined. 
If the baryon asymmetry we see today was present in
the thermal soup above the electroweak phase transition, in
the presence of the anomalous non-perturbative $B+L$-violating
electroweak interactions, then at least one lepton flavour must have been
conserved at the time, as we have briefly reviewed.
Whether or not an asymmetry remains in
the plasma is a physical question, which cannot depend on
how one chooses to define a ``lepton" or  ``lepton-number
violation".

We have therefore identified basis-independent combinations
of coupling constants that parametrize lepton-flavour
violation in the superpotential. The various couplings of
the Higgs and the leptons can be visualised as vectors
and planes in  a space spanned by the Higgs and lepton
superfields. The vectors $\mu$ and $\lambda_d^{pq}$ 
are directions that should be the Higgs, and
the $\lambda_e^i$ correspond to one or two
planes that should be spanned by the Higgs and
the $i$th lepton. It therefore takes three couplings
to construct a geometric quantity that measures
$R$ violation. One needs a 
Yukawa coupling, to provide a plane that would like to be spanned by
a Higgs and a lepton, and then one needs two further interactions:
one to choose a direction in the
plane to be the Higgs - the perpendicular direction in
the plane is now the lepton, and another to conflict
with this choice.  Our invariants correspond to the
projection of a vector that should be the Higgs onto
a direction chosen to be a lepton by other interactions.
This is a geometric, basis-independent  measure of
$R$ violation.

We have listed all the invariants that we can
construct out of three coupling constants, which is the minimum
required, as discussed above. It is geometrically clear that these 
invariants are not all independent: indeed, there are
321 of them with three generations of leptons. Therefore,
we have worked out how many independent invariants
there are, finding the 36 expected in the three-generation case.
In the two-generation case, we have also demonstrated explicitly 
how to express the remaining invariants
in terms of the independent ones.

In the early Universe, we can associate a lepton-flavour-violating rate
with each of the invariants.  This allows
us to discuss the survival of the baryon asymmetry of the Universe in
a basis-independent way. We have therefore determined
the bounds on our invariants that would allow a baryon
asymmetry to remain in the plasma before the electroweak phase
transition. These  bounds reduce to those of
~\cite{DE} if flavour dependence is neglected. In the 
basis-dependent formulation that we would
prefer to avoid, but which is more intuitive,
we find that a baryon asymmetry would remain
in the plasma if all the  dimensionless Yukawa
couplings violating  some
lepton flavour $i$  are less than $10^{-7}$,
and the superpotential mass term $\mu^i$ mixing
the Higgs and the $i$th lepton satisfies
$h_b \times \mu^i< 10^{-7} \mu^0$, where $h_b$ is
the bottom yukawa, and $\mu^0$ mixes the two Higges. 
These are the same bounds on the Yukawa couplings
as were found in previous work \cite{CDEO,DR},
but  the bound on $\mu^i$ is weaker.

As noted in the introduction, we have not included
the soft SUSY-breaking masses or the Higgs vacuum expectation
value in the construction
of our invariants. These would both be required for a complete
discussion of phenomenological bounds on
the $R$-violating couplings in our
basis-independent approach. It is at present unclear
to us whether there is any possible confusion that
could arise from the basis dependence of such
experimental bounds,  which are calculated in the
zero-temperature mass-eigenstate basis. We expect that
our invariants should be useful for comparing
bounds on $R$-violating couplings
calculated at different energy scales, e.g., the GUT scale, the
electroweak phase transition temperature, and at low energy, but such
an analysis is left to a future paper.

\subsection*{Acknowledgements}
One of us (S.D.) would like to thank Howie Haber for useful discussions.

\end{document}